

\font\twelverm=cmr10 scaled 1200    \font\twelvei=cmmi10 scaled 1200
\font\twelvesy=cmsy10 scaled 1200   \font\twelveex=cmex10 scaled 1200
\font\twelvebf=cmbx10 scaled 1200   \font\twelvesl=cmsl10 scaled 1200
\font\twelvett=cmtt10 scaled 1200   \font\twelveit=cmti10 scaled 1200
\font\twelvesc=cmcsc10 scaled 1200
\skewchar\twelvei='177   \skewchar\twelvesy='60
\def\twelvepoint{\normalbaselineskip=12.4pt
  \medskipamount=7.2pt plus2.4pt minus2.4pt
  \def\rm{\fam0\twelverm}          \def\it{\fam\itfam\twelveit}%
  \def\sl{\fam\slfam\twelvesl}     \def\bf{\fam\bffam\twelvebf}%
  \def\mit{\fam 1}                 \def\cal{\fam 2}%
  \def\tt{\twelvett}
  \def\sc{\twelvesc}
  \def\nullspace{\nulldelimiterspace=0pt \mathsurround=0pt }
  \def\big##1{{\hbox{$\left##1\vbox to 10.2pt{}\right.\nullspace$}}}
  \def\Big##1{{\hbox{$\left##1\vbox to 13.8pt{}\right.\nullspace$}}}
  \def\bigg##1{{\hbox{$\left##1\vbox to 17.4pt{}\right.\nullspace$}}}
  \def\Bigg##1{{\hbox{$\left##1\vbox to 21.0pt{}\right.\nullspace$}}}
  \textfont0=\twelverm   \scriptfont0=\tenrm   \scriptscriptfont0=\sevenrm
  \textfont1=\twelvei    \scriptfont1=\teni    \scriptscriptfont1=\seveni
  \textfont2=\twelvesy   \scriptfont2=\tensy   \scriptscriptfont2=\sevensy
  \textfont3=\twelveex   \scriptfont3=\twelveex  \scriptscriptfont3=\twelveex
  \textfont\itfam=\twelveit
  \textfont\slfam=\twelvesl
  \textfont\bffam=\twelvebf \scriptfont\bffam=\tenbf
  \scriptscriptfont\bffam=\sevenbf
  \normalbaselines\rm}

\def\beginlinemode{\endmode
  \begingroup\parskip=0pt \obeylines\def\\{\par}\def\endmode{\par\endgroup}}
\def\beginparmode{\endmode \begingroup \def\endmode{\par\endgroup}}
\let\endmode=\par
{\obeylines\gdef\
{}}
\def\singlespace{\baselineskip=\normalbaselineskip}
\def\oneandahalfspace{\baselineskip=\normalbaselineskip \multiply\baselineskip
     by 3 \divide\baselineskip by 2}
\def\doublespace{\baselineskip=\normalbaselineskip \multiply\baselineskip by 2}
\newcount\firstpageno \firstpageno=2
\footline={\ifnum\pageno<\firstpageno{\hfil}\else{\hfil\twelverm\folio\hfil}\fi}

\def\raggedcenter{\leftskip=4em plus 12em \rightskip=\leftskip
  \parindent=0pt \parfillskip=0pt \spaceskip=.3333em \xspaceskip=.5em
  \pretolerance=9999 \tolerance=9999 \hyphenpenalty=9999 \exhyphenpenalty=9999
}
\parskip=\medskipamount \twelvepoint \overfullrule=0pt
\def\title {\null\vskip 3pt plus 0.3fill \beginlinemode
   \doublespace \raggedcenter \bf}
\def\author{\vskip 3pt plus 0.3fill \beginparmode \raggedcenter \sc}

\def\abstract{\vskip 3pt plus 0.3fill \beginparmode
   \oneandahalfspace \narrower ABSTRACT:~~}
\def\endtitlepage{\vfill\eject\beginparmode}

\def\endit{\endmode\vfill\supereject\end}
\def\half{{\textstyle{ 1\over 2}}}

\def\sss{\scriptscriptstyle}
\def\gtwid{\mathrel{\raise.3ex\hbox{$>$\kern-.75em\lower1ex\hbox{$\sim$}}}}
\def\ltwid{\mathrel{\raise.3ex\hbox{$<$\kern-.75em\lower1ex\hbox{$\sim$}}}}

\def\Dt{\Delta T}
\def\s{\sigma}
\def\ahat{\hat a}
\def\bhat{\hat b}
\def\chat{\hat c}
\def\khat{\hat k}
\def\nhat{\hat n}
\def\cd{\!\cdot\!}
\def\wc{\widetilde C}
\def\kt{k_{\rm\sss T}}
\def\etal{et al.}
\def\la{\langle}
\def\ra{\rangle}
\def\frac#1#2{{\textstyle{#1\over#2}}}
\def\p{\varphi}

\singlespace
\rightline{UCSBTH--92--31}
\rightline{July 1992}

\doublespace
\title
THE ANGULAR DEPENDENCE OF THE THREE-POINT CORRELATION FUNCTION OF THE
COSMINC MICROWAVE BACKGROUND RADATION
AS PREDICTED BY INFLATIONARY COSMOLOGIES

\vskip 3pt plus 0.3fill \beginlinemode
\centerline{\sc Toby Falk, Raghavan~Rangarajan, and Mark Srednicki}

\vskip 3pt plus 0.1fill
\centerline{\rm Department of Physics, University of California,
                \ Santa Barbara, CA 93106}

\vskip 3pt plus 0.3fill \rm
\centerline{ABSTRACT}
\beginparmode \oneandahalfspace \narrower
Inflationary models predict a definite, model independent, angular dependence
for the three-point correlation function of $\Delta T/T$ at large angles
($\gtwid 1^\circ$) which we calculate.  The overall amplitude is model
dependent and generically unobservably small, but may be large in some specific
models.  We compare our results with other models of nongaussian fluctuations.

\noindent
{\it Subject headings:} cosmology: theory --- large-scale structure of universe

\endtitlepage
\baselineskip=17.5pt
Recent results from COBE (Smoot \etal\ 1992)
have provided strong evidence for fluctuations $\Dt$ in the cosmic microwave
background radiation~(CMBR) on large angular scales ($\gtwid 1^\circ$).
The two-point correlation function of $\Dt$
is consistent with a Harrison--Zel'dovich spectrum, as predicted
by inflationary cosmologies (Guth 1981).
Here we point out that inflation results in a definite and model
independent prediction for the
angular dependence of the three-point function as well.
The overall amplitude of the three-point function, however,
 is model dependent.
It is possible to construct models (discussed below) where this
amplitude would be large enough to be seen in the COBE data, but generically
it is expected to be too small to be observable.

For large angular scales, $\Dt$ is predicted to be (Peebles 1982;
Abbott \& Wise 1984; Bond \& Efstathiou 1987)
$$\Dt(\ahat) = \left({3\over5\pi}\right)^{1/2}Q\int d^3\!k\,\xi(\vec k)
               \, e^{i\vec k\cdot\ahat}\;,            \eqno(1)$$
where $\ahat$ is a unit vector on the sky,
$Q$ is the predicted RMS quadrupole amplitude, and the two-point correlation
function of the random variable $\xi(\vec k)$ is given by
$$\bigl\la \xi(\vec k_1)\xi(\vec k_2)\bigr\ra
                = k_1^{-3}\delta^3(\vec k_1+\vec k_2)\;,\eqno(2)$$
where $\delta^3(\vec k)$ is the three-dimensional Dirac delta function.
To study the two-point correlation function of $\Dt$ in a realistic setting,
we must first expand $\Dt$ in Legendre polynomials, remove the monopole,
dipole, and (following Smoot et al.~1992) quadrupole terms,
and weight the remaining terms to account for the finite beam width $\s$
of the antennas.  This results in (Bond \& Efstathiou 1987)
$$ e^{i\vec k\cdot\ahat} \to
    \sum_{l=3}^\infty i^l(2l+1)\,W_l\,j_l(k)\,P_l(\khat\cd\ahat)\;, \eqno(3)$$
where $k=|\vec k|$, $\khat=\vec k/k$, $j_l(k)$ is a spherical Bessel function,
and $W_l\simeq\exp[-\half(l+\half)^2\s^2]$;  Smoot \etal\ (1992) take
$\s=3.2^\circ$.  This implies that the two-point correlation function for $\Dt$
takes the form
$$\bigl\la \Dt(\ahat)\Dt(\bhat) \bigr\ra  = \frac65 Q^2 \, C_2(\ahat,\bhat)\;,
                                                                   \eqno(4)$$
where
$$C_2(\ahat,\bhat) = \sum_{l=3}^\infty (2l+1)\,C_l\,W_l^2\,
                                             P_l(\ahat\cd\bhat)\;,   \eqno(5)$$
with $C_l=2\int_0^\infty dk\,j_l^2(k)/k = 1/l(l+1)$.
For reference,
$C_2(\ahat,\bhat)$ is plotted as a function of the separation angle
$\alpha=\cos^{-1}(\ahat\cd\bhat)$ in Figure~(1).

In inflationary models, the random variable $\xi(\vec k)$ has its origin
in the fluctuations of a quantum field $\p$.
Working in momentum space, and assuming $\p$ is massless and noninteracting,
during the de~Sitter epoch the two-point correlation function of $\p$ is
$$\bigl\la \p(\vec k_1,\tau)\p(\vec k_2,\tau)\bigr\ra
                = 4\pi^3H^2 k_1^{-3}\delta^3(\vec k_1+\vec k_2)   \eqno(6)$$
in the long wavelength limit (Bunch \& Davies 1978).
Here $\tau$ is conformal time
and $H$ is the Hubble parameter, related to the (constant) scalar potential
$V_0$ via $H^2=8\pi V_0/3$.  (We take $\hbar=c=G=1$.)
The three-point correlation function vanishes identically.
However,
if the field $\p$ has any sort of interaction~(with itself or with other
fields), nontrivial higher-point correlation functions will appear,
and these will be passed on to the temperature perturbations
(Allen, Grinstein, \& Wise 1987).  The simplest possible interaction,
and one which will generically be present, is a modification of
the scalar potential from a constant to $V(\p)=V_0+\frac16\mu\p^3$.
To leading order in $\mu$, this interaction
implies a three-point function which can be calculated via standard
methods of quantum field theory in curved space (Birrell \& Davies 1982).
In the long wavelength limit, we find
$$\eqalignno{
\bigl\la \p(\vec k_1,\tau)\p(\vec k_2,\tau)\p(\vec k_3,\tau)\bigr\ra &=
        \frac23\pi^3\mu H^2\, (k_1 k_2 k_3)^{-3}\, F(k_1,k_2,k_3) \cr
&\qquad\qquad\qquad \times  \delta^3(\vec k_1+\vec k_2+\vec k_3)\;,&(7)\cr}$$
where
$$\eqalignno{
F(k_1,k_2,k_3) &=  (k_1^3+k_2^3+k_3^3)\bigl[\log(\kt|\tau|)+\gamma\bigr]\cr
&\qquad\qquad -(k_1^2+k_2^2+k_3^2)\kt + k_1 k_2 k_3\;.  &(8)\cr}$$
Here $k_i=|\vec k_i|$ and $\kt=k_1+k_2+k_3$; $\gamma$ is Euler's constant.
The factor of $\log(\kt|\tau|)$ in equation~(8)
is minus the number of e-foldings
from the time at which fluctuations of wavenumber~$\kt$ first passed outside
the horizon until the end of inflation.  For the relevant length scales,
one typically finds $\log(\kt|\tau|) \simeq -60$
(e.g., Bardeen, Steinhardt, \& Turner 1982),
and so the term with this factor dominates over the others.
{}From here on we will approximate $F(k_1,k_2,k_3)$ as
$$F(k_1,k_2,k_3) \simeq -\beta\,(k_1^3+k_2^3+k_3^3)     \eqno(9)$$
and treat $\beta\simeq 60$ as independent of $k_i$.
A similar logarithmic dependence on $k_i$ (with a different origin)
is also present in equation~(2), and is also routinely
ignored.

The fluctuations in $\p$ become fluctuations in the CMBR in one of two ways.
If the energy stored in $\p$ is eventually converted to ordinary matter
and energy (including the CMBR), then the fluctuations are adiabatic,
and the correlation functions of $\xi(\vec k)$ are the same (up to a sign)
as those of $\p(\vec k)/(4\pi^3H^2)^{1/2}$.  In this case we typically find
$\mu/H\ltwid 10^{-7}$ (Hodges \etal\ 1991), but larger values may be possible
if we fine-tune the scalar potential.
If, instead, the energy stored in $\p$ is eventually converted to
dark matter, then the fluctuations are isocurvature (Linde 1984),
and the correlation functions of
$\xi(\vec k)$ are again the same (up to a sign) as those of
$\p(\vec k)/(4\pi^3H^2)^{1/2}$ (Allen \etal\ 1987).
In this case there are no obvious restrictions on $\mu/H$.

Neglecting, for the moment,
the effects of finite beam width and the need to subtract the
$l=0,1,2$ terms in $\Dt$, we can combine Eqs.~(1), (2), (6), (7), and (9)
to compute an ``unsubtracted'' three-point correlation function for $\Dt$.
We find
$$\bigl\la \Dt(\ahat)\Dt(\bhat)\Dt(\chat) \bigr\ra_{\rm unsub}
         =  \varepsilon\,{1\over\pi}\left({3\over5}\right)^{3/2}
            \beta\,{\mu\over H}\,Q^3\,\wc_3(\ahat,\bhat,\chat)\;, \eqno(10)$$
where $\varepsilon=\pm 1$ is a model dependent sign, and
$$\eqalignno{
\wc_3(\ahat,\bhat,\chat) &=
       {1\over 3(2\pi)^2}
       \int d^3\!k_1\,d^3\!k_2\,d^3\!k_3\,
       \delta^3(\vec k_1+\vec k_2+\vec k_3) \cr
 &\qquad\qquad\qquad\times  (k_1^{-3}k_2^{-3}+\ldots)
       \,e^{i(\vec k_1\cdot\ahat +
              \vec k_2\cdot\bhat +
              \vec k_3\cdot\chat)}  \cr
\noalign{\medskip}
    &=
       {1\over 3(2\pi)^2}
       \int d^3\!k_1\,k_1^{-3}
       \,e^{i\vec k_1\cdot(\ahat-\chat)}
                           \int d^3\!k_2\,k_2^{-3}
       \,e^{i\vec k_2\cdot(\bhat-\chat)} + \ldots \cr
\noalign{\medskip}
   &= \frac13\left[\wc_2(\ahat,\chat)\wc_2(\bhat,\chat) +
                   \wc_2(\bhat,\ahat)\wc_2(\chat,\ahat) +
                   \wc_2(\chat,\bhat)\wc_2(\ahat,\bhat)\right]\;.\cr
                                                     &{} &(11)\cr}$$
In the last line, we have defined an ``unsubtracted'' dimensionless
two-point correlation function
$$\eqalignno{
\wc_2(\ahat,\bhat) &= {1\over 2\pi} \int d^3\!k\,k^{-3}
                      \,e^{i\vec k\cdot(\ahat-\bhat)} \cr
\noalign{\medskip}
                   &= \sum_{l=0}^\infty (2l+1)\,C_l\,
                      P_l(\ahat\cd\bhat)\;. &(12)\cr}$$
Equation~(11) is a beautifully simple formula, but clearly is only a formal
relation, since the monopole term in $\wc_2$ is infinite.
Equation~(11) must be corrected to remove the $l=0,1,2$ terms, and to account
for finite beam width.  This is straightforward, and ultimately yields
$$\bigl\la \Dt(\ahat)\Dt(\bhat)\Dt(\chat) \bigr\ra
         =  \varepsilon\,{1\over\pi}\left({3\over5}\right)^{3/2}
            \beta\,{\mu\over H}\,Q^3\,C_3(\ahat,\bhat,\chat)\;, \eqno(13)$$
where
$$\eqalignno{
C_3(\ahat,\bhat,\chat)
&=\frac13\sum_{j,k,l=3}^\infty (2j+1)(2k+1)(2l+1)\,(C_j C_k +
                                                    C_k C_l +
                                                    C_l C_j)\cr
&\qquad\qquad\qquad  \times W_j W_k W_l\,
                     f_{jkl}(\ahat,\bhat,\chat)\;, &(14)\cr}$$
and we have defined
$$f_{jkl}(\ahat,\bhat,\chat)={1\over4\pi}\int d\Omega_n\,P_j(\ahat\cd\nhat)
                                                         P_k(\bhat\cd\nhat)
                                                         P_l(\chat\cd\nhat)\;.
                                                                \eqno(15)$$
Here $d\Omega_n$ denotes integration over the unit vector $\nhat$.
If desired, $f_{jkl}(\ahat,\bhat,\chat)$
can be expanded in spherical harmonics of each of
the three unit vectors by making use of standard relations
[for example, eqs.~(9.79) and (16.90) of Merzbacher (1970)],
but the resulting formula for the coefficients
is unwieldy and will not be presented here.

$C_3(\ahat,\bhat,\chat)$ is plotted in Figures~(2) and~(3)
for an equilateral triangle on the sky:
$\ahat\cd\bhat=\bhat\cd\chat=\chat\cd\ahat=\cos\alpha$.
Surprisingly, for this case
it is given to a good approximation by $[C_2(\ahat,\bhat)]^2$,
mimicking the relation between the unsubtracted functions
$\wc_2$ and $\wc_3$.  Clearly to check for the presence of
a nonzero three-point function in the COBE data, one should look at
the smallest feasible values of $\alpha$.
If a signal is found, the specific prediction of equation~(14) could
be checked by looking at other configurations of $\ahat$, $\bhat$, and
$\chat$, such as all three in a line on the sky.

The nonzero three-point function implies that the fluctuations are
nongaussian.  The particular form of the three-point function specified
by equation~(9) also arises (to leading order in $\lambda$) in models in
which fluctuations in the gravitational potential $\phi$ are given by a
local, nonlinear transformation of a gaussian random field $\xi$:
$\phi(\vec x)=\xi(\vec x)+\lambda\xi(\vec x)^2+\ldots\,$.
Such models have been investigated by Kofman \etal\ (1991),
Moscardini \etal\ (1991), and Scherrer (1992).
In particular, Kofman \etal\ (1991) have noted that such effects can arise
due to nonlinearities in the classical evolution equations for
certain inflationary models, even if the initial fluctuations are gaussian.
Other nongaussian models (which do not result from inflation)
give different three-point functions.
For example, Weinberg \& Cole (1992) have studied
models in which the density fluctuations $\delta\rho/\rho$
are given by a local,
nonlinear transformation of a gaussian random field; this yields
$F\propto (k_1k_2k_3)^2(k_1^{-1}+k_2^{-1}+k_3^{-1})$ to leading order
in $\lambda$.  Still another type of model has been proposed by
Scherrer \& Schaefer (1992); it gives $F\propto (k_1k_2k_3)^{3/2}$.
These results for $F$ imply that, if we normalize
the deviations from gaussian statistics at small scales,
inflationary models will produce larger deviations
on large scales than these alternatives.
We would argue that future work on nongaussian models should focus
on those which reproduce the
three-point function reported here, since it is a generic prediction
of inflationary cosmologies.

We would like to thank Mark Wise for pointing out to us both the feasibility
and potential interest of this calculation, and for providing us a copy
of his unpublished notes on the related calculations reported by
Allen et al.  This work was supported in part by NSF Grant
No.~PHY-91-16964.

\vfill\eject
\centerline{REFERENCES}
\vskip0.3in

\frenchspacing
\hoffset=0truein
\hsize=6.25truein
\parindent=-0.25truein

Abbott, L. F., \& Wise, M. B. 1984, Phys. Lett., B135, 179

Allen, T. J., Grinstein, B., \& Wise, M. B. 1987, Phys. Lett., B197, 66

Bardeen, J. M., Steinhardt, P. J., \& Turner, M. S. 1983, Phys.~Rev.~D,
28, 679

Birrell, N. D., \& Davies, P. C. W. 1982,
Quantum Fields in Curved Space (Cambridge: Cambridge University Press)

Bond, J. R., \& Efstathiou, G. 1987, MNRAS, 226, 655

Bunch, T. S., \& Davies, P. C. W. 1978, Proc. R. Soc. Lon., A360, 117

Guth, A. 1981, Phys.~Rev.~D, 23, 347

Hodges, H.~M., Blumenthal, G.~R., Kofman, L.~A., \& Primack, J.~R.
1990, Nucl. Phys., B335, 197

Kofman, L., Blumenthal, G.~R., Hodges, H.~M., \& Primack, J.~R. 1991,
in Large Scale Structures and Peculiar Motions in the Universe,
ed. D. W. Latham \& L. A. Nicolaci da Costa (San Francisco: Astron.
Soc. Pac.)

Linde, A. D. 1984, JETP Lett., 40, 1333

Merzbacher, E. 1970, Quantum Mechanics (New York: Wiley)

Moscardini, L., Matarrese, S., Lucchin, F., \& Messina, A. 1991,
MNRAS, 248, 424

Peebles, P. J. E. 1982, ApJ, 263, L1

Smoot, G. F., \etal\ 1992, ApJ, in press

Scherrer, R. J. 1992, ApJ, 390, 330

Scherrer, R. J., \& Schaefer, R. K. 1992, Ohio State Univ. preprint
OSU-TA-7/92

Weinberg, D. H., \& Cole, S. 1991, U. C. Berkeley preprint CfPA-TH-91-025

\vfill\eject
\centerline{FIGURE CAPTIONS}
\vskip0.3in

Fig.~1.---The dimensionless two-point correlation $C_2(\ahat,\bhat)$
as a function of~$\alpha$, where $\ahat\cd\bhat=\cos\alpha$,
for $\sigma=3.2^\circ$.
({\sl Description: $C_2(\alpha)$ is the same as the center of the gray band in
Fig.~3 of Smoot et al.\ but with a different vertical scale; $C_2(0)=3.04$.})

Fig.~2.---The dimensionless three-point correlation
$C_3(\ahat,\bhat,\chat)$ as a function of~$\alpha$, where
$\ahat\cd\bhat= \bhat\cd\chat= \chat\cd\ahat= \cos\alpha$,
for $\sigma=3.2^\circ$.  ({\sl Description: $C_3(\alpha)$ is defined only for
$\alpha$ between $0^\circ$ and $120^\circ$.  Over this range it looks a lot
like $C_2(\alpha)^2$; $C_3(0)=9.15$ and $C_3(120^\circ)=0$.})

Fig.~3.---Same as Figure~2, with the vertical scale expanded to
show the large angle structure.

\endit
\end